\documentclass[showpacs,aps,prl,twocolumn]{revtex4-1}

\usepackage{graphicx}
\usepackage{amsmath}
\usepackage{subfigure}
\usepackage{dcolumn}

\begin{document}

\textheight 9.32in

%%%%%%%%%%%%%%%%%%%%%%%%%%%%%%%%%%%%%%%%%%%%%%%%%%%%%%%%%%%%%%%%%%%%%%%%%%%%%%%%

\title{First-principles investigation of morphotropic transitions and
  phase-change functional responses in BiFeO$_3$--BiCoO$_3$
  multiferroic solid solutions}

\author{Oswaldo Di\'eguez}
\author{Jorge \'I\~niguez}

\affiliation{Institut de Ci\`encia de Materials de Barcelona (ICMAB-CSIC),
             Campus UAB, 08193 Bellaterra, Spain}

%%%%%%%%%%%%%%%%%%%%%%%%%%%%%%%%%%%%%%%%%%%%%%%%%%%%%%%%%%%%%%%%%%%%%%%%%%%%%%%%

\begin{abstract}
  We present an {\em ab initio} study of the BFCO solid solution
  formed by multiferroics BiFeO$_3$ (BFO) and BiCoO$_3$ (BCO). We find
  that BFCO presents a strongly discontinuous morphotropic transition
  between BFO-like and BCO-like ferroelectric phases. Further, for all
  compositions such phases remain (meta)stable and retain
  well-differentiated properties. Our results thus suggest that an
  electric field can be used to switch between these structures, and
  show that such a switching involves large {\em phase-change} effects
  of various types, including piezoelectric, electric, and
  magnetoelectric ones.
\end{abstract}

%\date{\today}

\pacs{77.84.-s, 75.85.+t, 71.15.Mb}

% 77. Dielectrics, piezoelectrics, and ferroelectrics and their properties
% 77.80.-e Ferroelectricity and antiferroelectricity
% 77.84.-s Dielectric, piezoelectric, ferroelectric, and antiferroelectric materials

% 75. Magnetic properties and materials
% 75.80.+q Magnetomechanical and magnetoelectric effects,
% magnetostriction
% 75.85.+t Magnetoelectric effects, multiferroics

% 71.  Electronic structure of bulk materials
% 71.15.Mb Density functional theory, local density approximation,
% gradient and other corrections 

\maketitle

%%%%%%%%%%%%%%%%%%%%%%%%%%%%%%%%%%%%%%%%%%%%%%%%%%%%%%%%%%%%%%%%%%%%%%%%%%%%%%%%

Functional oxides attract attention because of their potential for
designing materials tailored for specific applications. A lot of work
focuses on BiFeO$_3$ (BFO), one of the few compounds that is
magnetoelectric (ME) multiferroic -- i.e., displays coupled electric
and magnetic orders -- at room
temperature~\cite{Catalan2009AM}. Interest in BFO has been recently
refueled by the discovery that an electric field $\boldsymbol{\cal E}$
can be used to switch between two different ferroelectric (FE) phases
of epitaxially-compressed films~\cite{Zeches2009S}. Such a
$\boldsymbol{\cal E}$-switching has a number of functional effects
associated to it, as the phases involved are markedly dissimilar in
terms of cell shape (the switching thus implies a large piezoelectric
effect) and magnetism (ME effect). Hence, BFO films offer the
appealing possibility of obtaining {\em phase-change} functional
responses of various kinds. Here we propose that some BFO-based solid
solutions are ideally suited to this end, and present illustrative
first-principles results for BiFe$_{1-x}$Co$_{x}$O$_3$.

{\sl Materials-design aspects}.-- The $\boldsymbol{\cal E}$-switching
in BFO films involves two phases~\cite{hatt10}: one that is similar to
the rhombohedral structure of bulk BFO and has a polarization
$\boldsymbol{P}$ roughly along the [111] pseudo-cubic direction ($R$
phase in the following); and a phase with a unit cell of very large
aspect ratio ($c/a \sim$~1.25) and $\boldsymbol{P}$ roughly parallel
to [001] ({\em super-tetragonal} or $T$ phase). First-principles work
has shown that these phases revert their relative stability as a
function of epitaxial strain~\cite{hatt10, Wojdel2010PRL}, and that
there is a strain range in which both can
exist~\cite{Wojdel2010PRL}. It has also been predicted that, even in
absence of stabilizing fields, BFO presents many $T$ phases that are
local energy minima~\cite{dieguez11}. The theory is thus compatible
with the observation that $\boldsymbol{\cal E}$ fields can be used to
switch between different FE phases of BFO.

These results suggest that, to find materials in which the
$\boldsymbol{\cal E}$-switching is possible, one must look for
compounds displaying a strongly discontinuous transition between two
FE phases; further, the FE phases should be robustly stable and their
polarizations point along markedly different directions, so that it is
easy to switch between them by applying properly oriented fields. An
obvious strategy is to look for chemical substitutions of BFO that may
result in a morphotropic phase boundary (MPB) between the $R$ phase of
the pure compound and a second FE structure. Among many possibilities,
the BiFe$_{1-x}$Co$_{x}$O$_3$ (BFCO) solid solution seems particularly
promising. Note that bulk BiCoO$_3$ (BCO) is a ME multiferroic that
presents a super-tetragonal FE phase~\cite{Belik2006CM}; thus, BFCO is
likely to display a $R$-$T$ morphotropic transition analogous to the
one induced by epitaxial compression in BFO films.

The X-ray experiments of Azuma {\em et al}.~\cite{Azuma2008JJAP}
confirmed that, as the Co content grows, BFCO moves from $R$ to $T$
traversing a narrow region of presumably monoclinic ($M$)
symmetry. (Similar results have been obtained for thin
films~\cite{Yasui2009}.)  This is reminiscent of what occurs in
prototype piezoelectric PbZr$_{1-x}$Ti$_x$O$_3$ (PZT), where the $M$
phase found at the MPB \cite{Noheda1999APL} is characterized by its
{\em structural softness} and large electromechanical
responses~\cite{bellaiche00}. However, as far as we know, no enhanced
response has been observed in BFCO, which questions the existence of a
PZT-like $M$ phase in this compound~\cite{miura10}. Note that this is
encouraging in the present context: For $\boldsymbol{\cal
  E}$-switching purposes, we would like BFCO's $R$ and $T$ phases to
be relatively stiff (as opposed to soft) and stable.

{\sl BFCO's morphotropic transitions}.-- We used the so-called
``GGA+$U$'' approach to density functional theory (DFT) as implemented
in the {\sc VASP} package~\cite{vasp}, the calculation details being
essentially identical as in previous studies of similar
materials~\cite{fn:simulations}. We worked with the 40-atom cell
depicted in the insets of Fig.~1(b), which allows us to describe the
$R$ and $T$ phases of interest~\cite{dieguez11} and vary the ratio of
Co atoms $x$ in steps of $1/8$. Figure~1(a) shows the formation energy
of the phases investigated as a function of composition; this is
defined as $E_{\rm f} = E - (1-x) E_{\rm BFO} - x E_{\rm BCO}$, where
$E$ is the energy of a particular BFCO structure of composition $x$,
and $E_{\rm BFO}$ and $E_{\rm BCO}$ are the ground-state energies of
the pure compounds. Figures~1(b) and 1(c-d) show, respectively, the
results for the polarization (only for selected structures) and cell
parameters.

The six ponts at $x=0$ in Fig.~1(a) correspond to the stable BFO
phases described in Ref.~\onlinecite{dieguez11}, whose properties are
summarized in Table~\ref{tab_phases}. The ground state is the
well-known FE phase of the compound; we call it $R$-G, noting its
G-type anti-ferromagnetic (AFM) spin order (i.e., nearest-neighboring
irons have anti-parallel spins). We also considered four FE $T$-phases
with C-AFM order (i.e., parallel spins along the stretched lattice
vector and anti-parallel in-plane). We label these phases according to
their symmetry, which is not tetragonal for any of them: We have an
orthorhombic phase ($T_{\rm ort}$), two monoclinic phases of type
$M_A$ (i.e., with $\boldsymbol{P}$ in the $(1\bar{1}0)$
plane~\cite{vanderbilt01}; we call them $T_A$), and a monoclinic $M_C$
phase (i.e., with $\boldsymbol{P}$ in the $(100)$ plane; we call it
$T_C$). Finally, we also considered the paraelectric phase $p$-G.

\begin{table}
\setlength{\extrarowheight}{1mm}
\caption{Stable phases considered for BFO and BCO. We show the label
  for each phase (see text), space group (S.~G.), polarization
  magnitude ($\mu$C/cm$^2$) and angle (relative to the perpendicular
  to the plane defined by the two shortest pseudo-cubic lattice
  vectors for the $T$ phases, and to the pseudo-cubic $[111]$ for the $R$
  phases), and energy (meV/f.u.) above that of the most stable
  phase.}
\vskip 1mm
\begin{tabular}{ccccc}
\hline\hline
Material  &  Phase  &  S. G.  &  Polarization  &  $\Delta E$  \\
\hline
BFO  &  $T_A$-C          &  $Pc$      &  120 (23$^\circ$)  &  106  \\
     &  $T_C$-C          &  $Cm$      &  150 (20$^\circ$)  &  103  \\
     &  $T_{\rm ort}$-C  &  $Pna2_1$  &  139  (0$^\circ$)  &   99  \\
     &  $T_A$-C          &  $Cc$      &  145 (19$^\circ$)  &   96  \\
     &  $p$-G            &  $Pnma$    &                 0  &   27  \\
     &  $R$-G            &  $R3c$     &   91  (0$^\circ$)  &    0  \\
\hline
BCO  &  $R_A$-G          &  $Pc$      &   82  (5$^\circ$)  &   52  \\
     &  $p$-G            &  $Pnma$    &                 0  &   51  \\
     &  $T$-C            &  $P4mm$    &  167  (0$^\circ$)  &    0  \\
\hline\hline
\end{tabular}
\label{tab_phases}
\end{table}

\begin{figure}
\centering
\includegraphics[height=52mm]{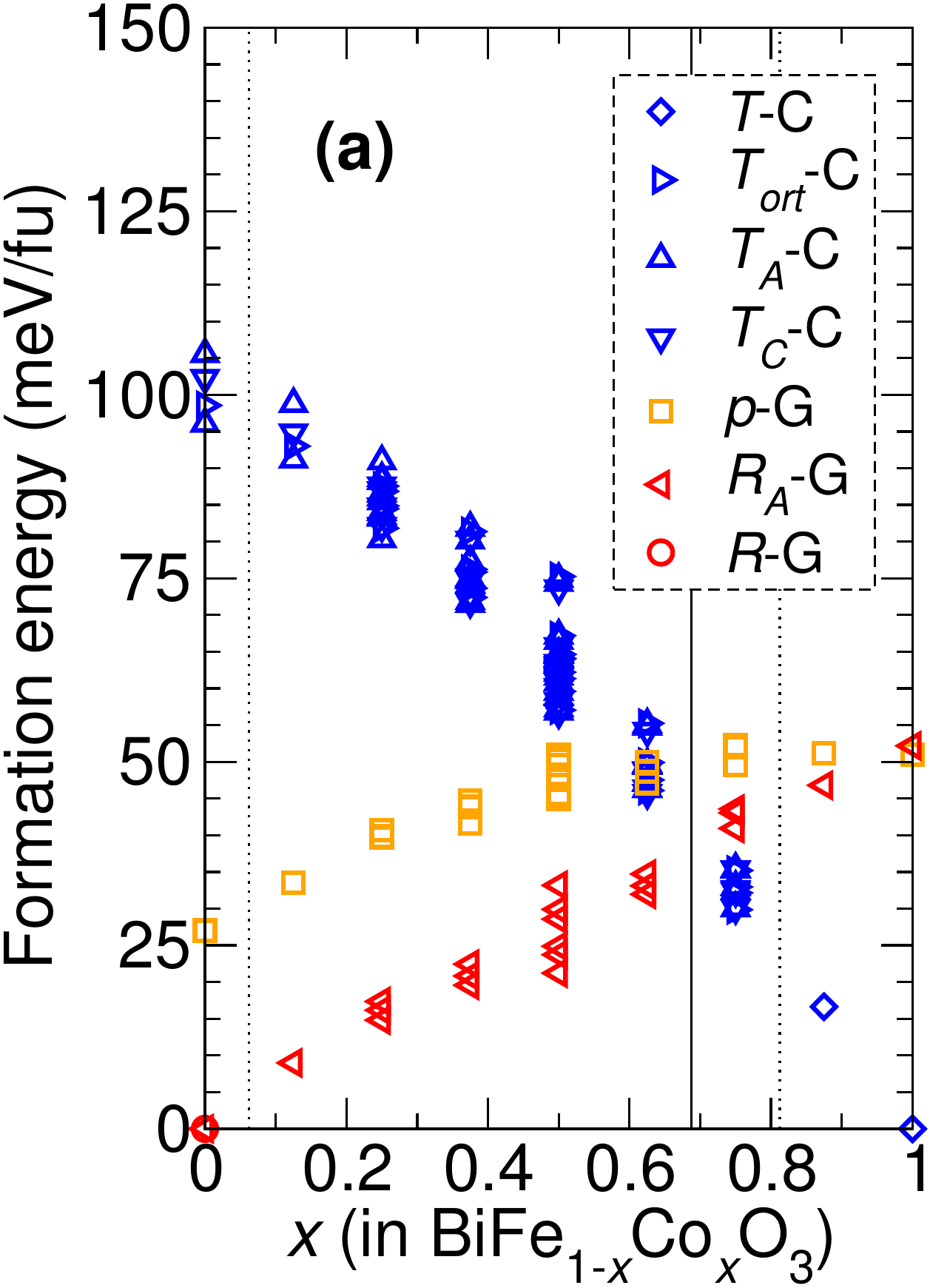}
\hspace{2mm}
\includegraphics[height=52mm]{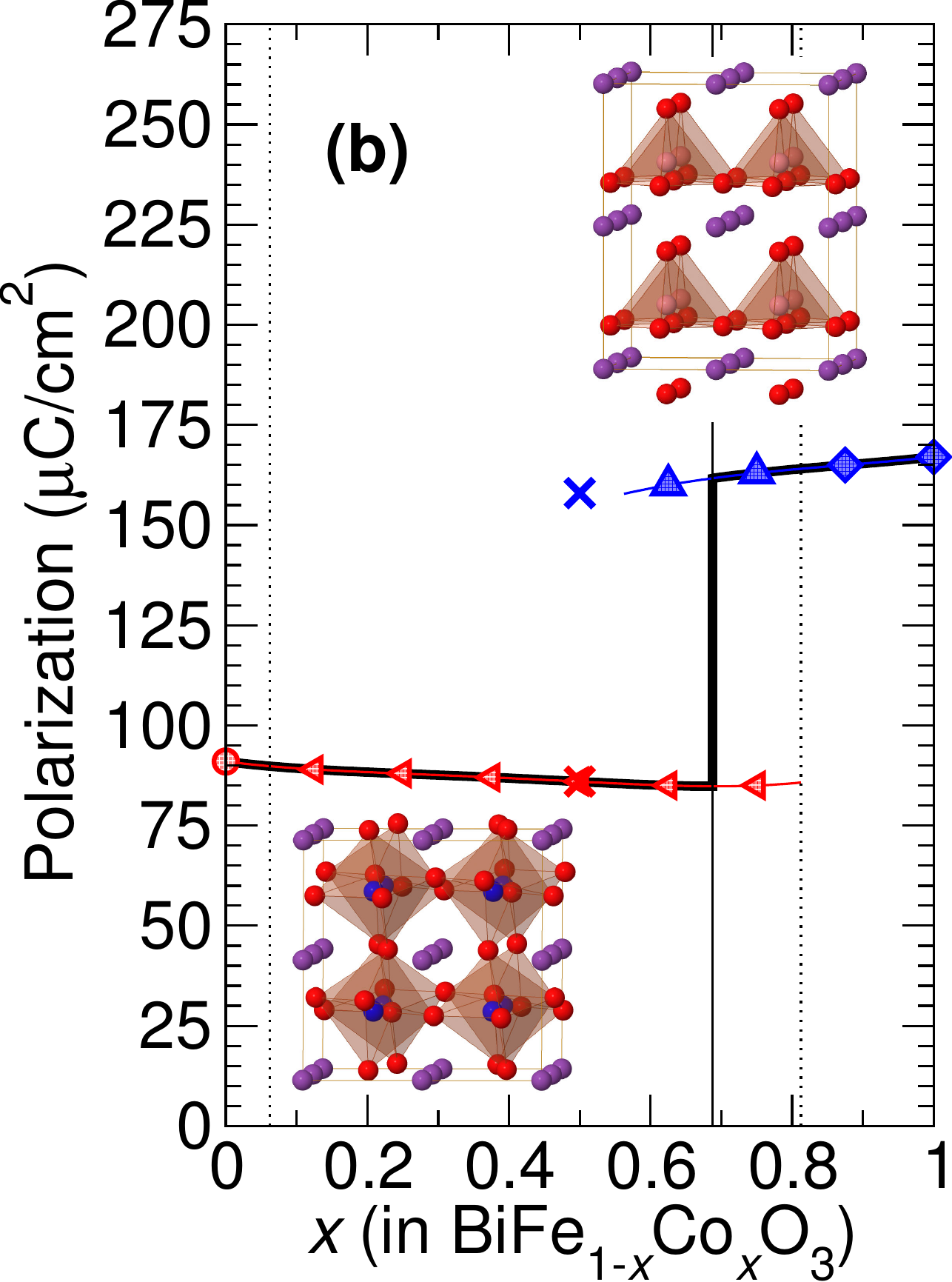}
\\
\vspace{3mm}
\includegraphics[width=80mm]{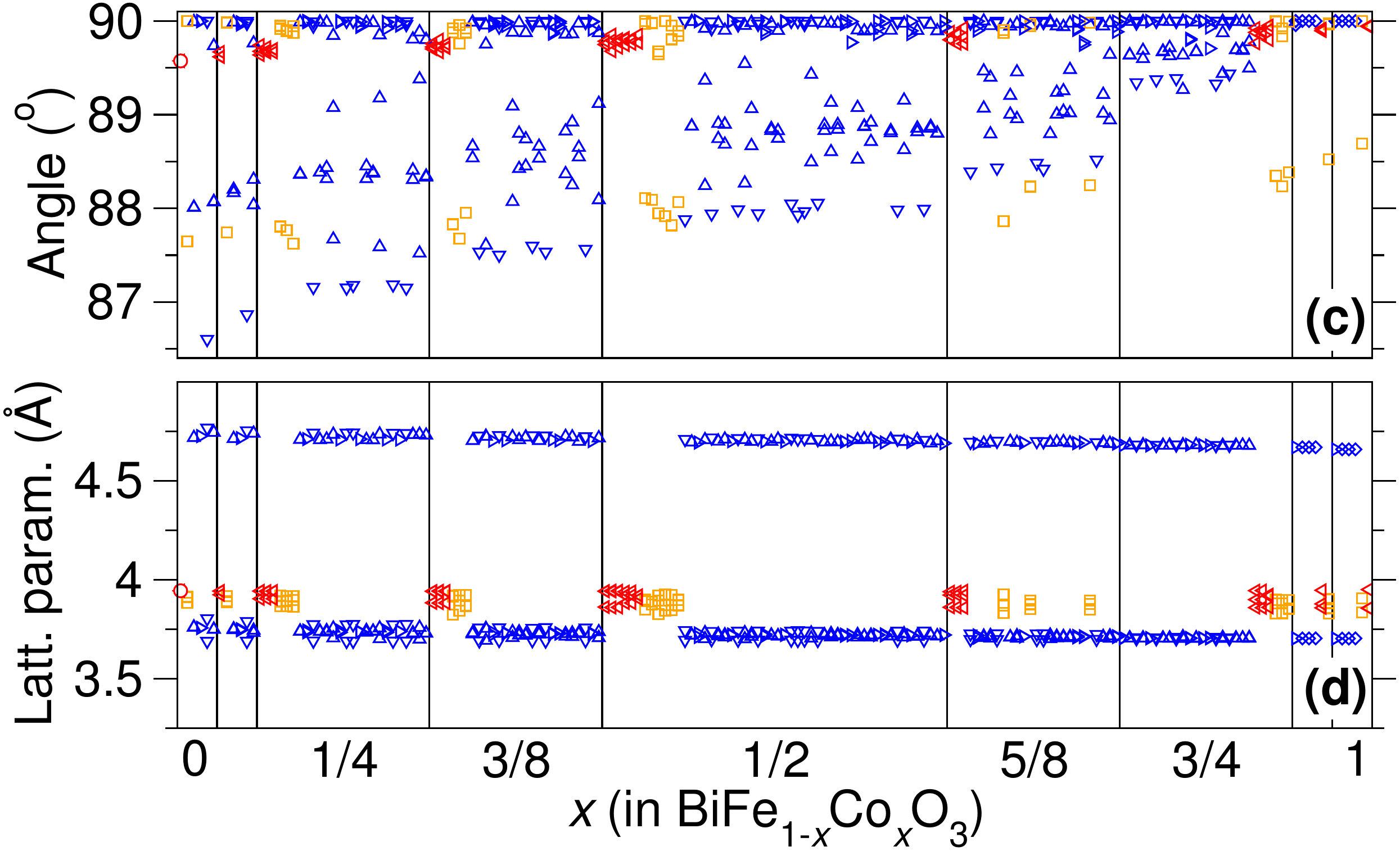}
\caption{(Color online.) (a)~Formation energy {\em vs} composition for
  all studied phases. Labels as in Table~\protect\ref{tab_phases};
  various instances of the same symbol at a given $x$ correspond to
  different Fe/Co arrangements. (b)~Polarization magnitude for the
  most stable structures (joined by thick line); a few others shown
  for comparison; crosses correspond to rocksalt-ordered $T_A$ and
  $R_A$ structures. Insets: BFO ($R$-G) and BCO ($T$-C)
  structures. (c)~Pseudo-cubic lattice constants (lower panel) and
  angles (upper panel) of the structures in panel~(a); same-$x$
  configurations given in order of increasing energy. Dotted vertical
  lines mark second-order $R$-$R$ and $T$-$T$ transitions (see text);
  solid line marks the first-order $R$-$T$ transition.}
\label{fig_structures}
\end{figure}

To search for BCO phases, we substituted irons by cobalts in the six
structures just described; then, for each of them we ran a short
molecular dynamics (with random initial velocities to break all
symmetries) followed by a full relaxation. We thus obtained the three
solutions listed in Table~I. All four $T$ structures relaxed to the
$T$-C phase known to be BCO's ground state~\cite{Belik2006CM}; our
computed structure ($a = b =$~3.70~\AA\ and $c/a =$~1.26) is in
reasonable agreement with experiment ($a = b =$~3.729~\AA\ and $c/a =$
1.267~\cite{Belik2006CM}) and previous DFT results~\cite{Cai2007JCP,
  Ravindran2008AM}. The optimization starting from BFO's $R$-G phase
led to a structure in which the $R3c$ symmetry is slightly broken to
monoclinic $M_A$ [faces of O$_6$ octahedra lying on (111) planes form
isosceles, instead of equilateral, triangles]; we call it
$R_A$-G. Finally, BCO's $p$-G phase is analogous to BFO's. For
intermediate compositions, we considered BFO's stable phases and
studied all the inequivalent Fe/Co arrangements at the perovskite
B-sites. In all cases we ran a molecular dynamics followed by a full
structural relaxation; the result of each such optimization renders a
data point in Fig.~\ref{fig_structures}(a).

Many conclusions can be drawn from Fig.~\ref{fig_structures}. Most
importantly, we found that BFCO undergoes a strongly discontinuous
transition between a $R$ phase ($c/a \sim$~1) and a $T$ phase ($c/a
\sim$~1.25) at $x \approx$~0.7. We also found that the $R$ and $T$
phases (as well as the $p$ phase) are stable for all
compositions~\cite{fn:stability}, a feature likely related to Bi's
peculiar bonding properties~\cite{dieguez11}. Note also that the $R$
and $T$ phases retain their main features -- i.e., values of
polarization and structural parameters -- for all considered
compositions and Fe/Co arrangements. Hence, according to our results,
BFCO may allow for an $\boldsymbol{\cal E}$-controlled switching
between two distinct FE phases in a wide composition range. Further,
the robustly stable character observed for the $R$ and $T$ phases
suggests there may be a relatively large experimental freedom to tune
BFCO (e.g., by varying the composition or epitaxial conditions of a
film) and optimize the switching.

As regards the structure of the $R$ phases, Figs.~1(b-d) show no major
changes occur when moving from BFO's $R$-G to BCO's
$R_A$-G. Nevertheless, we identified a distinct effect: For $x>0$ all
the $R$ structures present the distortion of the oxygen-octahedron
faces described above for pure BCO. By inspecting the cases in which
the Fe/Co arrangement is compatible with the rhombohedral 3-fold axis,
we found this symmetry breaking renders a $M_A$ structure, as in pure
BCO. We thus label these phases as $R_A$-G in
Fig.~\ref{fig_structures}, where a dotted line at $x \approx$~0.07
marks a continuous transition between $R$-G and $R_A$-G.

As for the $T$ phases, Figs.~1(c-d) show they evolve continuously as
the Co content increases, and become truly tetragonal (i.e., the
pseudo-cubic lattice constants $a$ and $b$ become equal, and the
angles turn 90$^{\circ}$) for $x \approx$~0.8; this second-order
transition to BCO's $T$-C phase is marked by a dotted line in Fig.~1.

Our results thus confirm the $R$-$T$ MPB reported by Azuma {\em et
  al}.~\cite{Azuma2008JJAP}; yet, the agreement between theory and
experiment is far from perfect. At a quantitative level, our
calculations place the MPB at $x \approx$~0.7, at variance with the
value of $x \approx$~0.35 obtained by extrapolating the data of
Ref.~\onlinecite{Azuma2008JJAP} to low temperatures. We think this
deviation can be partly related to DFT's limitations to predict
accurately the relative stablity of the $R$ and $T$ phases, as
recently discussed for BFO~\cite{dieguez11}. The simulations, on the
other hand, seem reliable when they predict that BFCO displays no
PZT-like $M$ phase acting as a {\em bridge} between the $R$ and $T$
structures at the MPB; rather, our results would be compatible with a
$R$-$T$ phase coexistence at the MPB region. Finally, more structural
measurements are needed to confirm the monoclinic symmetry predicted
for the $R$ and $T$ phases at intermediate compositions.

{\sl BFCO's phase-change properties}.-- The $R$-$T$ switching involves
a large change in BFCO's unit cell, which results in a phase-change
{\em piezoelectric} effect. It also involves a large change in
polarization: We may switch between $P_z \approx$~53~$\mu$C/cm$^2$
($R$ phase) and $P_z \approx$~165~$\mu$C/cm$^2$ ($T$ phase), which may
prove useful in the design of field-effect and other
devices. Additionally, the $R$ and $T$ phases present very different
dielectric and piezoelectric responses: For example, for $x = 1/2$ and
a rocksalt Fe/Co order, we obtained an approximately diagonal static
dielectric tensor for the $R_A$-G phase, with $\epsilon_{xx}^{\rm
  latt} \approx$~46; in contrast, the analogous $T_C$-C phase presents
an anisotropic response with $\epsilon_{xx}^{\rm latt} \approx$~259,
$\epsilon_{yy}^{\rm latt} \approx$~122, and $\epsilon_{zz}^{\rm latt}
\approx$~16. (We observed similar trends for piezoelectricity. These
results reflect a well-known fact for FE perovskite oxides: making
$\boldsymbol{P}$ rotate is energetically less costly than changing its
magnitude~\cite{bellaiche00}.) Hence, the $R$-$T$ switching also
allows for a large dielectric and piezoelectric
tunability~\cite{fn:gap}.

The $R$-$T$ switching also involves a change in the spin order, which
moves from G-AFM to C-AFM. If any, the net magnetization of such AFM
structures will be a small one arising from spin canting (see
Ref.~\onlinecite{Wojdel2010PRL} for representative results for BFO);
thus, the associated phase-change effect will be tiny. On the other
hand, work on BFO~\cite{hatt10,dieguez11} shows that the magnitude of
the exchange interactions varies considerably between the $R$ and $T$
phases. We found that such a differentiated behavior also occurs in
BFCO, resulting in markedly different N\`eel temperatures ($T_{\rm
  N}$'s) and response properties.

\begin{figure}
\centering
\includegraphics[width=50mm, angle=-90]{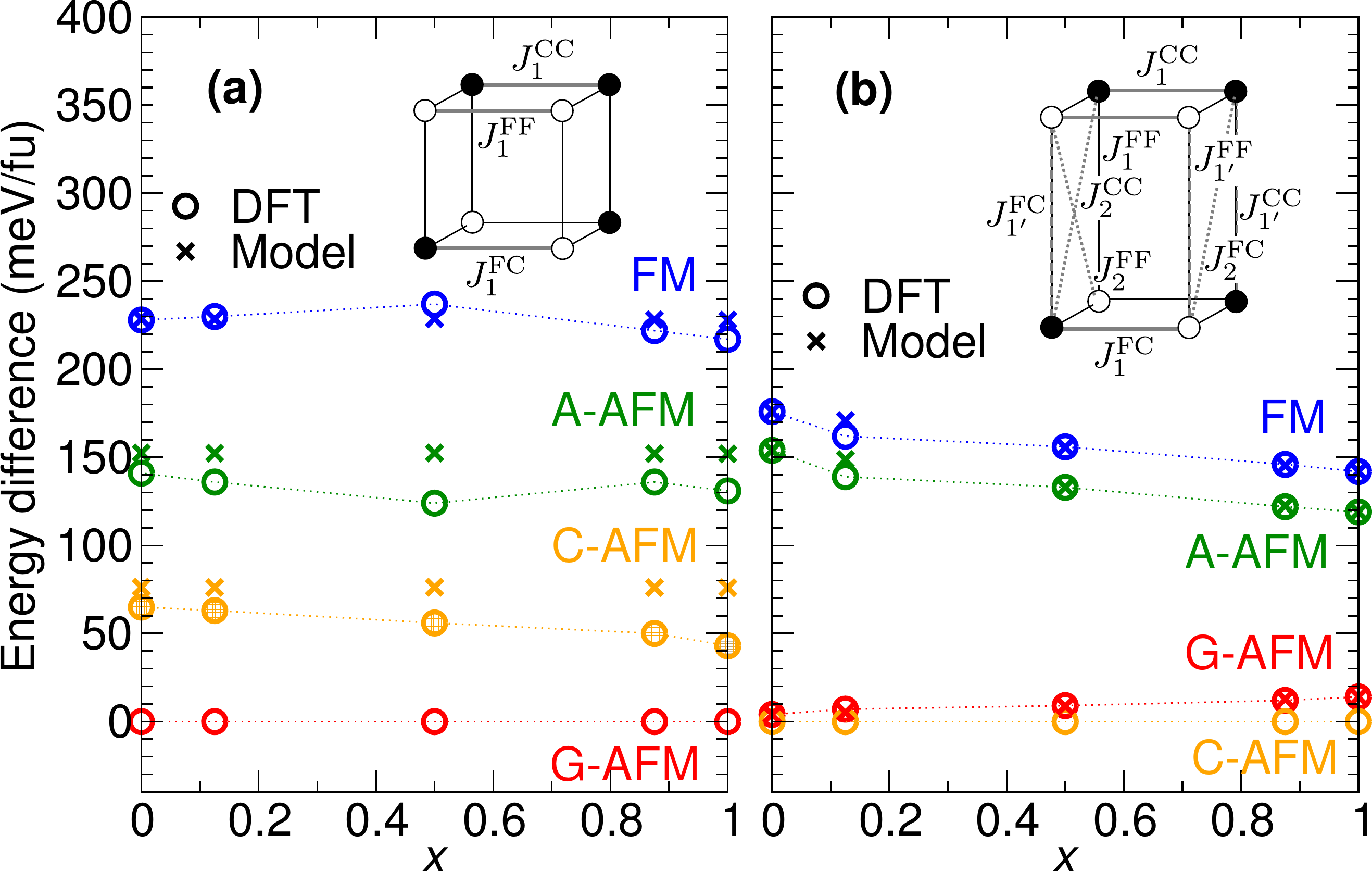}
\\
\vspace{3mm}
\includegraphics[height=50mm]{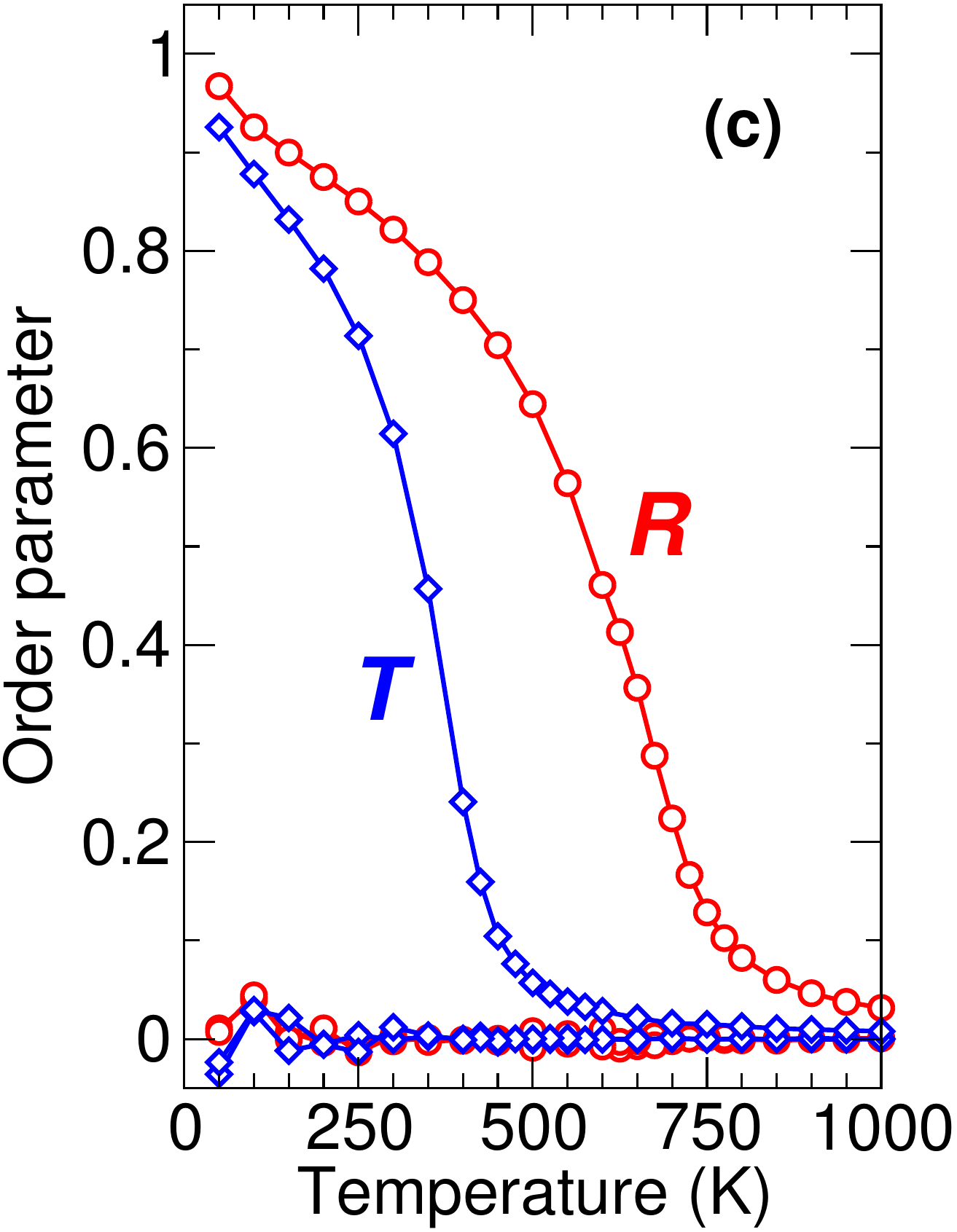}
\hspace{2mm}
\includegraphics[height=50mm]{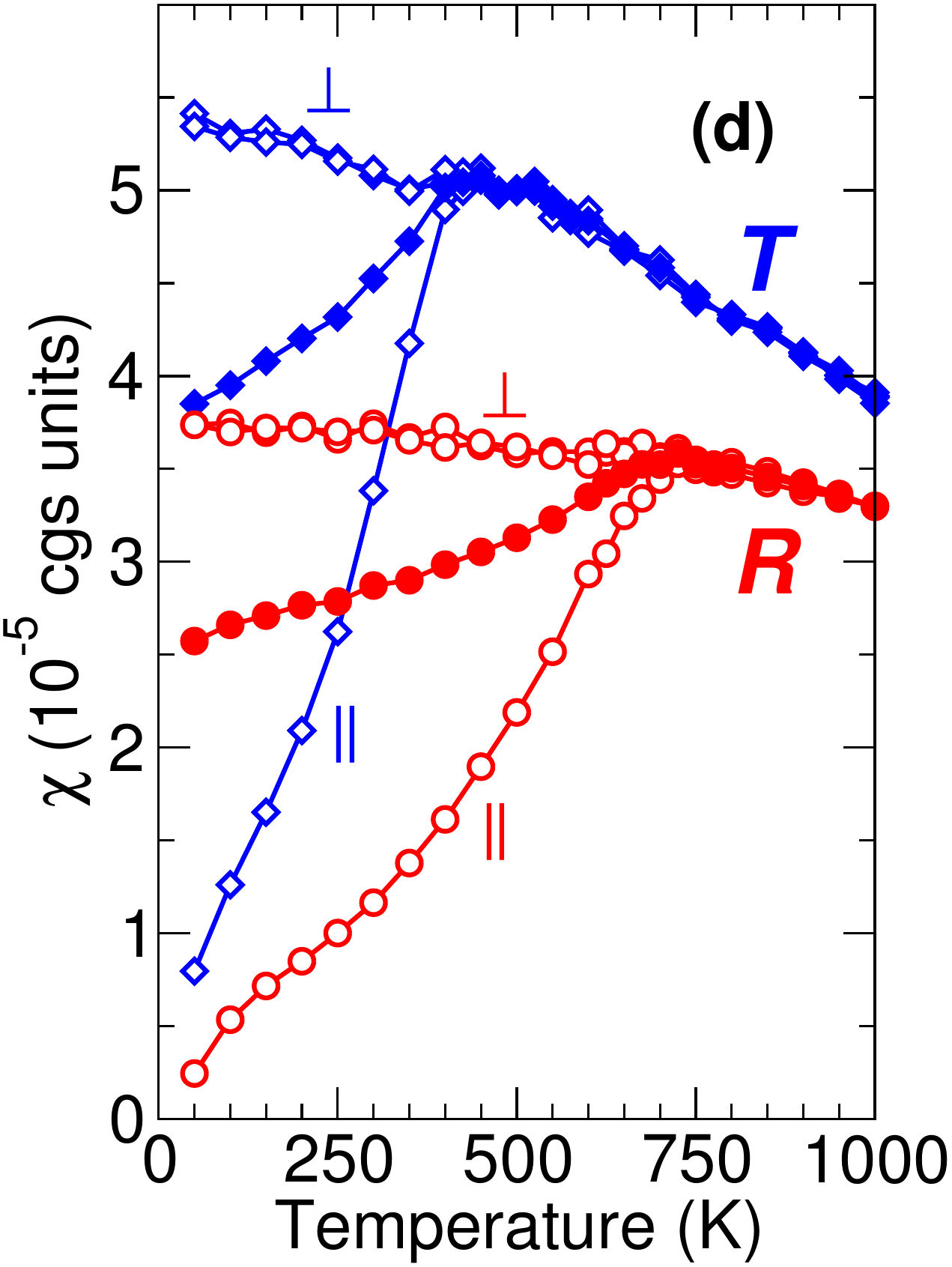}
\caption{(Color online.)  (a) and (b):~Energy differences between spin
  orders for representative BFCO structures (see text); $R$ phases in
  (a) and $T$ phases in (b). Circles show DFT results; crosses
  correspond to model Hamiltonians including the $J$ couplings defined
  in insets (see text). (c)~Temperature dependence of AFM order
  parameter (G-type for $R$ and C-type for $T$).  (d)~Temperature
  dependence of the total magnetic susceptibility (solid symbols), and
  of its components [parallel ($\parallel$) and perpendicular
  ($\perp$); open symbols] with respect to the direction of the AFM
  order parameter.}
\label{fig_magmom}
\end{figure}

To investigate BFCO's magnetic properties, we considered the most
stable $R$ and $T$ configurations at $x=$~0, $1/8$, $7/8$, and 1, as
well as the lowest-energy structures with a rocksalt Fe/Co arrangement
at $x = 1/2$~\cite{fn:ferri}. We computed the energies of the
following spin orders: ferromagnetic (FM), C-AFM, G-AFM, and A-AFM (as
obtained from G-AFM by making spins parallel in-plane); the results
are shown in Figs.~2(a) and 2(b). Most notably, we found that the
structure type determines the hierarchy of spin arrangements:
Independently of the Co content, all the $R$ phases have the G-AFM
order as the most stable one, followed by C-AFM, A-AFM, and FM; in
turn, C-AFM is the ground state of all the $T$ phases, followed by
G-AFM, A-AFM, and FM. From a model Hamiltonian perspective, this
implies that the magnetic couplings will depend strongly on the atomic
structure ($R$ or $T$) and weakly on the chemical details (Fe/Co ratio
and spatial arrangement). Indeed, we were able to capture the essence
of our DFT results in the two simple Heisenberg models, one for {\em
  all} the $R$ phases and one for {\em all} the $T$ phases, described
in the following.

We write the energy of a spin configuration as $E = E^0 + 1/2
\sum_{ij} J_{ij} \, \boldsymbol{S}_i \cdot \boldsymbol{S}_j$, where
$E^0$ is a reference energy, $J_{ij}$ is the exchange coupling between
Fe/Co atoms $i$ and $j$, and $\boldsymbol{S}_i$ is the
three-dimensional spin of atom $i$. For simplicity, here we assume
$|\boldsymbol{S}_i| =$~1; the reported $J_{ij}$ values have been
calculated accordingly.

In the $R$ phases, any Fe/Co atom has six Fe/Co first nearest
neighbors (fnn's) that are roughly equivalent (they are equivalent if
the symmetry is exactly rhombohedral). Thus, we can tentatively take
$J_{ij} = J_1$ as our single fnn interaction, and also neglect further
couplings. From our results for the pure compounds, we obtained
$J_1^{\rm FF} = 38.0$~meV and $J_1^{\rm CC} = 38.1$~meV, and from the
results at $x = 1/2$ we got $J_1^{\rm FC} = 38.1$~meV. The minimal
Hamiltonian thus defined predicts the energies shown as crosses in
Fig.~2(a); it clearly captures the essential magnetic interactions in
BFCO's $R$ phases.

For the $T$ phases, we take advantage of the (approximate) tetragonal
symmetry and consider only two fnn couplings: in-plane $J_{1}$ and
out-of-plane $J_{1'}$. Note that by choosing $J_{1} > 0$ and $J_{1'} <
0$ we would correctly reproduce the C-AFM ground state, but wrongly
predict the A-AFM order as the least favorable one. Hence, we need to
introduce a new interaction $J_{2}$ between second nearest neighbors
(snn) [see inset in Fig.~2(b)]. With this model we obtained the
results shown as crosses in Fig.~2(b); the computed parameters are (in
meV): $J_{1}^{\rm CC} = 30.9$, $J_{1}^{\rm FC} = 35.0$, $J_{1}^{\rm
  FF} = 40.8$, $J_{1'}^{\rm CC} = 2.3$, $J_{1'}^{\rm FC} = 3.3$,
$J_{1'}^{\rm FF} = 4.5$, $J_{2}^{\rm CC} = 2.3$, $J_{2}^{\rm FC} =
2.0$, and $J_{2}^{\rm FF} = 1.6$. Note that all the computed $J_{1'}$
and $J_{2}$ constants are positive, implying that all the out-of-plane
interations are anti-ferromagnetic in the $T$ phases of
BFCO. Ultimately, the $J_{2}$ couplings prevail and render the
out-of-plane parallel spin alignment of the C-AFM ground state.

We solved these Hamiltonians using standard Monte Carlo
techniques. Figures~2(c) and 2(d) show representative results obtained
for $x=1/2$ systems with Fe/Co atoms randomly distributed in a
10$\times$10$\times$10 periodically-repeated simulation box. We found
that the $R$ and $T$ cases present markedly different $T_{\rm N}$'s of
about 675~K and 425~K, respectively. These results are compatible with
the experimental values for BFO (643~K)~\cite{Catalan2009AM} and BCO
(470~K)~\cite{Belik2006CM}, and reflect the weak out-of-plane
interactions that result in a lower $T_{\rm N}$ in the $T$
case. Figure~2(d) shows the magnetic susceptibility: At room
temperature, the total susceptibility of the $T$ phase almost doubles
the result for the $R$ phase; hence, we predict that the $R$-$T$
switching will have a clear magnetic signature.

We have thus shown that the BiFe$_{1-x}$Co$_x$O$_3$ solid solutions
are likely to present large electric-field-driven {\em phase-change}
effects of various types (piezoelectric, electric,
magnetoelectric). We hope our predictions will attract interest
towards such promising materials and effects.

Supported by MICINN-Spain [Grants No. MAT2010-18113,
  No. MAT2010-10093-E, and No. CSD2007-00041; {\em Ram\'on y Cajal}
  program (OD)]. Computing facilities provided by RES and
CESGA. Discussions with M.~Bibes, J.~Fontcuberta, and F.~S\'anchez are
acknowledged.

%%%%%%%%%%%%%%%%%%%%%%%%%%%%%%%%%%%%%%%%%%%%%%%%%%%%%%%%%%%%%%%%%%%%%%%%%%%%%%%%

%%%%%%%%%%%%%%%%%%%%%%%%%%%%%%%%%%%%%%%%%%%%%%%%%%%%%%%%%%%%%%%%%%%%%%%%%%%%%%%%


\begin{thebibliography}{99}

\bibitem{Catalan2009AM} G.\ Catalan and J.\ F.\ Scott, Adv.\ Mater.\
  {\bf 21}, 2463 (2009).

\bibitem{Zeches2009S} R.\ J.\ Zeches {\em et al}., Science {\bf 326},
  977 (2009).

\bibitem{hatt10} A.J.~Hatt, N.A.~Spaldin, and C.~Ederer, Phys. Rev. B
  {\bf 81}, 054109 (2010).

\bibitem{Wojdel2010PRL} J.C.\ Wojde\l ~and J.\ \'I\~niguez, Phys.\
  Rev.\ Lett.\ {\bf 105}, (2010).

\bibitem{dieguez11} O.~Di\'eguez, O.E.~Gonz\'alez-V\'azquez,
  J.C.~Wojde\l, and J.~\'I\~niguez, Phys. Rev. B {\bf 83}, 094105
  (2011).

\bibitem{Belik2006CM} A.A.\ Belik {\em et al}., Chem.\ Mater.\ {\bf
  18}, 798 (2006).

\bibitem{Azuma2008JJAP} M.\ Azuma {\em et al}., Jpn.\ J.\ Appl.\
  Phys.\ {\bf 47}, 7579 (2008).

\bibitem{Yasui2009} S.\ Yasui {\em et al}., Appl.\ Phys.\ {\bf 105},
  061620 (2009); S.\ Yasui {\em et al.}, Jpn.\ J.\ Appl.\ Phys.\ {\bf
    48}, 09KD06 (2009).

\bibitem{Noheda1999APL} B.\ Noheda {\em et al}., Appl.\ Phys.\ Lett.\
  {\bf 74}, 2059 (1999).

\bibitem{bellaiche00} L.~Bellaiche, A.~Garc\'{\i}a, and D.~Vanderbilt,
  Phys. Rev. Lett. {\bf 84}, 5427 (2000); H.~Fu and R.E.~Cohen, Nature
  {\bf 403}, 281 (2000).

\bibitem{miura10} K.~Miura, M.~Kubota, M.~Azuma, and H.~Funakubo,
  Jpn. J. Appl. Phys. {\bf 49}, 09ME07 (2010).

\bibitem{vasp} G.\ Kresse and J.\ Furthmuller, Phys.\ Rev.\ B {\bf
    54}, 11169 (1996); G.\ Kresse and D.\ Joubert, Phys.\ Rev.\ B {\bf
    59}, 1758 (1999).

\bibitem{fn:simulations} We used the so-called ``PBEsol''
  approximation [J.P.~Perdew {\em et al}., Phys.\ Rev.\ Lett.\ {\bf
    100}, 136406 (2008)], which describes well the relative stability
  of BFO's $R$ and $T$ phases~\protect\cite{dieguez11}. We used the
  ``Hubbard~$U$'' correction of Dudarev {\em et al}. [Phys.\ Rev.\ B
  {\bf 57}, 1505 (1998)], with $U_{\rm eff}$ values of 4~eV and 6~eV
  for the 3$d$ electrons of Fe and Co, respectively; these choices,
  which are common in the literature, affect the composition at which
  the MPB is obtained, but not our qualitative results. We used the
  ``projector augmented wave'' method [P.E.~Blochl, Phys.\ Rev.\ B
  {\bf 50}, 17953 (1994)] to represent the ionic cores, solving for
  the following electrons: Fe's 3$p$, 3$d$, and 4$s$; Co's 3$d$, and
  4$s$; Bi's 5$d$, 6$s$, and 6$p$; and O's 2$s$ and 2$p$. Other
  details as in~\protect\cite{dieguez11}.

\bibitem{vanderbilt01} We follow the notation for the $M$ phases
  introduce by Vanderbilt and Cohen [Phys. Rev. B {\bf 63}, 094108
    (2001)].

\bibitem{Cai2007JCP} M.-Q. Cai {\em et al}., J. Chem. Phys. {\bf 126},
  154708 (2007).

\bibitem{Ravindran2008AM} P. Ravindran, R. Vidya, O. Eriksson, and
  H. Fjellv{\aa}g, Adv. Mater. 20, 1353 (2008).

\bibitem{fn:stability} Our structural optimizations were designed to
  render energy minima; we confirmed the stability of selected
  structures. We estimated the energy barrier separating $R$ and $T$
  phases near the MPB, obtaining values ($\sim$~100~meV/f.u)
  comparable with the barrier associated to the Landau limit for FE
  switching in typical materials.

\bibitem{fn:gap} The computed electronic band gaps present a large
  spread of about 1~eV; we did not observe any distinct band-gap
  change associated to the $R$-$T$ switching.

\bibitem{fn:ferri} At $x =$~1/2, an ordered Fe/Co arrangement might
  result in {\em ferrimagnetism} (e.g., if a G-AFM spin structure is
  combined with rocksalt Fe/Co order). However, we obtained a very
  weak dependence of BFCO's energy on the Fe/Co arrangement; thus,
  such ordered structures seem unlikely.

\end{thebibliography}
\end{document}